\documentclass[3p,times,procedia]{elsarticle}
\usepackage{nupha_ecrc}

\volume{00}

\firstpage{1}

\journalname{Nuclear Physics A}

\runauth{}

\jid{nupha}

\jnltitlelogo{Nuclear Physics A}

\usepackage{amssymb}
\usepackage{lineno}
\usepackage{amsmath}
\usepackage{color}
\usepackage{graphicx}
\usepackage{wrapfig}
\usepackage{multirow}
\usepackage{subfigure}
\usepackage{lineno}
\usepackage[rightcaption]{sidecap}

\biboptions{compress}

\usepackage[figuresright]{rotating}

\newcommand {\dg}   {\Delta\gamma}

\newcommand {\psiPP}    {\psi_{\rm PP}}
\newcommand {\psiSP}    {\psi_{\rm SP}}

\newcommand {\psiTPC}    {\psi_{\rm TPC}}
\newcommand {\psiZDC}    {\psi_{\rm ZDC}}

\begin{document}

\begin{frontmatter}



\dochead{XXVIIIth International Conference on Ultrarelativistic Nucleus-Nucleus Collisions\\ (Quark Matter 2019)}

\title{Search for CME in U+U and Au+Au collisions in STAR with different approaches of handling backgrounds}


\author{Jie Zhao (for the STAR collaboration)}

\address{Department of Physics and Astronomy, Purdue University, West Lafayette, IN, 47907, USA}

\begin{abstract}
	The chiral magnetic effect (CME) refers to charge separation along a strong magnetic field between left- and right-handed quarks, caused by interactions with topological gluon fields from QCD vacuum fluctuations. We present two approaches to handle the dominant elliptic flow ($v_2$) background in the three-particle correlator ($\Delta\gamma_{112}$), sensitive to CME.

	In the first approach, we present the $\Delta\gamma_{112}$ and $\Delta\gamma_{123}$ measurements in U+U and Au+Au collisions. While hydrodynamic simulations including resonance decays and local charge conservation predict that $\Delta\gamma_{112}$ scaled by $N_{\rm part}/v_2$ will be similar in U+U and Au+Au collisions, the projected B-field exhibits a distinct difference between the two systems and with varying $N_{\rm part}$. Therefore, U+U and Au+Au collisions 
provide configurations with different expectations for both CME signal and background.
Moreover, the three-particle observable $\Delta\gamma_{123}$ scaled by $N_{\rm part}/v_3$ provide baseline measurement for only the background. 

	In the second approach, we handle the $v_2$ background by measuring $\Delta\gamma_{112}$ with respect to the planes of spectators 
measured by Zero Degree Calorimeters and participants measured by Time Projection Chamber.
These measurements contain different amounts of contributions from CME signal (along B-field, due to spectators) and $v_2$ background (determined by the participant geometry). 
With the two $\Delta\gamma_{112}$ measurements, the possible CME signal and the background contribution can be determined. 
We report such a measurement in Au+Au collisions at $\sqrt{s_{NN}}=$ 27 GeV with the newly installed event plane detector, 
and report the new findings in U+U system where the spectator-participant plane correlations are expected to differ from those in Au+Au collisions.
\end{abstract}


\end{frontmatter}



\section{Introduction}

Quark interactions with fluctuating topological gluon field can induce chirality imbalance and local parity violation in quantum chromodynamics (QCD)~\cite{Lee:1974ma,Kharzeev:1998kz,Kharzeev:1999cz}.
This can lead to electric charge separation in the presence of a strong magnetic field ($B$),
a phenomenon known as the chiral magnetic effect (CME)~\cite{Fukushima:2008xe,Muller:2010jd}.
Such a $B$-field may present in non-central heavy-ion collisions, generated by the spectator protons at early times~\cite{Kharzeev:2007jp,Asakawa:2010bu}.
Extensive theoretical and experimental efforts have been devoted to the search for the CME-induced charge separation along $B$ in heavy-ion collisions~\cite{Kharzeev:2015znc,Zhao:2018ixy,zhao:225,Zhao:2019hta}.

\section{Results}
We present two approaches to handle the dominant elliptic flow ($v_2$) background in the observable $\dg_{112}$ (charge separation across second-order event plane), which is sensitive to CME.

\begin{figure}[htbp!]
    \centering
    \includegraphics[width=6.0cm]{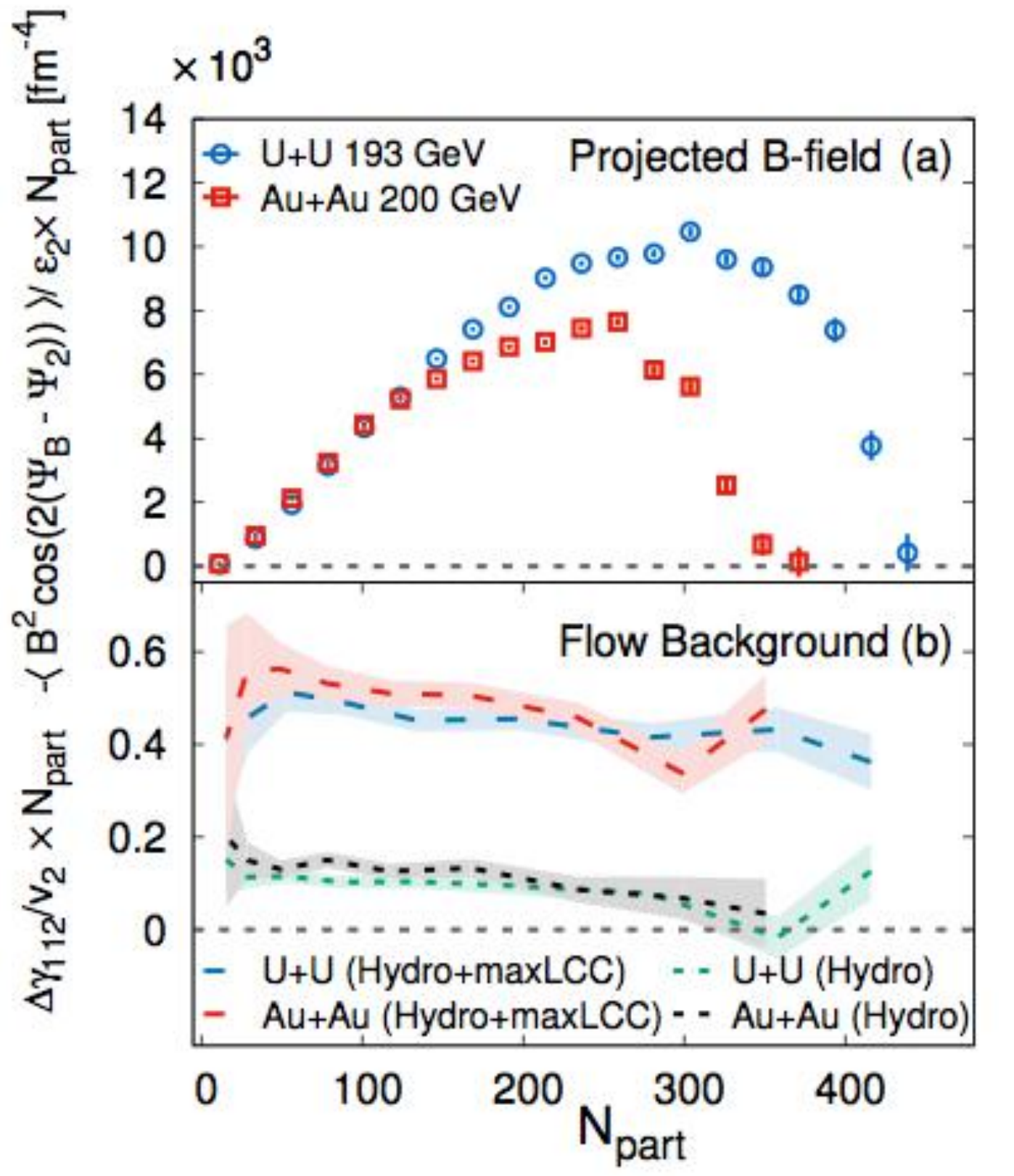}
    \includegraphics[width=6.0cm]{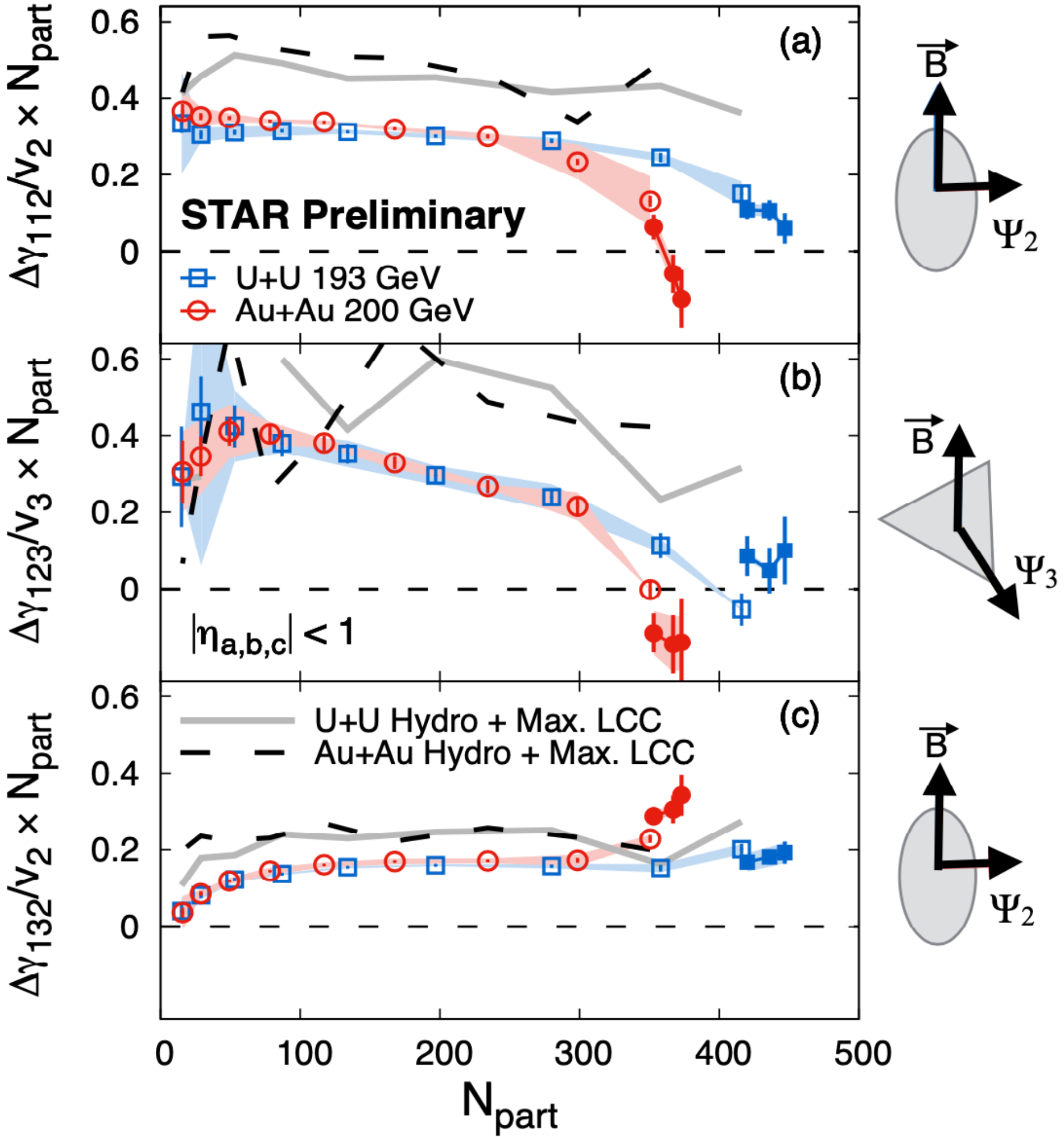}
    \caption{
		(Left upper panel) Predictions from MC-Glauber model~\cite{Chatterjee:2014sea} for projected magnetic field in Au+Au and U+U collisions. 
		(Left lower panel) Predictions for flow driven background using hydrodynamic simulations~\cite{Schenke:2019ruo}. 
		(Right) The $\Delta\gamma_{112}$, $\Delta\gamma_{123}$ and $\Delta\gamma_{132}$ measurements in U+U and Au+Au collisions.
    }
    \label{fig1}
\end{figure}

In the first approach, we present the $\Delta\gamma_{112}$, $\Delta\gamma_{123}$, and $\Delta\gamma_{132}$ measurements in U+U and Au+Au collisions.
The systematic studies of the $\Delta\gamma_{112}$, $\Delta\gamma_{123}$, and $\Delta\gamma_{132}$ in those two systems can provide insights on the CME signal and background behaviors. 
Left top panel of Fig.~\ref{fig1} show the expected B-field from MC-Glauber calculations~\cite{Chatterjee:2014sea}, which indicate that U+U and Au+Au have large B-field difference at large $N_{\rm part}$.
Charge separation driven by CME should be sensitive to such difference.
On the other hand, background model studies using hydrodynamic simulations~\cite{Schenke:2019ruo} indicate background to be similar between U+U and Au+Au as seen in Fig.~\ref{fig1} (left bottom panel). 
Furthermore, the third-harmonic event plane ($\psi_{3}$) is not expected to be correlated  with the magnetic field. 
Thus, one does not expect CME contribute to $\Delta\gamma_{123}$.

Right panels in Fig.~\ref{fig1} show the $\Delta\gamma_{112}$, $\Delta\gamma_{123}$, and $\Delta\gamma_{132}$ measurements in U+U and Au+Au collisions.
Background contributions based on hydrodynamic simulations with local charge conservation and global momentum conservation are included for comparison.
The mixed-harmonic correlations do not follow signal-only or background-only expectations.
Interesting features in ultra-central collisions are observed, which need further investigations.

In the second approach, we study the $\dg$ measurements with respect to the participant plane ($\psiPP$) and spectator plane ($\psiSP$).
The CME refers to charge separation along the strong magnetic field. 
The magnetic field is mainly produced by spectator protons in heavy-ion collisions,
strongest perpendicular to the $\psiSP$.
On the other hand, the major elliptic flow background is determined by the participant geometry,
largest in the $\psiPP$.
The $\psiSP$ and the $\psiPP$ can be assessed, experimentally in STAR,
	by the spectator neutrons in Zero Degree Calorimeters ($\psiZDC$) and by midrapidity particles in the Time Projection Chamber ($\psiTPC$), respectively.
The $\dg$ measurements with respect to $\psiZDC$ and $\psiTPC$
can therefore resolve the possible CME signal (and the background).
Consider the measured $\Delta\gamma$ to be composed of the $v_2$ background and the CME signal:
\begin{linenomath}
\begin{equation}
    \begin{split}
        \rm \dg\{\psiTPC\} = \dg_{CME}\{\psiTPC\} + \dg_{Bkg}\{\psiTPC\}, \ \dg\{\psiZDC\} = \dg_{CME}\{\psiZDC\} + \dg_{Bkg}\{\psiZDC\}.
    \end{split}
    \label{eqA}
\end{equation}
\end{linenomath}
Assuming the CME is proportional to the magnetic field squared and background is proportional to $v_{2}$~\cite{Xu:2017qfs},
both projected onto the $\psi$ direction, we have:
\begin{linenomath}
\begin{equation}
    \begin{split}
    \rm \dg_{CME}\{\psiTPC\} = \it{a}\rm \dg_{CME}\{\psiZDC\}, \ \dg_{Bkg}\{\psiZDC\}=\it{a}\rm \dg_{Bkg}\{\psiTPC\},  \\
    \end{split}
    \label{eqB}
\end{equation}
\end{linenomath}
where $a=\langle \rm cos2(\psiZDC-\psiTPC)\rangle$.
The parameter $a$ can be readily obtained from the $v_2$ measurements:
\begin{linenomath}
\begin{equation}
    \it a= v_{2}\{\rm \psiZDC\}/\it v_{2}\{\rm \psiTPC\}.
    \label{eqC}
\end{equation}
\end{linenomath}
The CME signal relative to the inclusive $\Delta\gamma\{\rm \psiTPC\}$ measurement is then given by:
\begin{linenomath}
\begin{equation}
    \begin{split}
		\rm f^{EP}_{CME} = \dg_{CME}\{\psiTPC\}/\dg\{\psiTPC\} = \it (A/a-1)(1/a^{2}-1),
    \end{split}
    \label{eqD}
\end{equation}
\end{linenomath}
where:
\begin{linenomath}
\begin{equation}
    A=\dg\{\psiZDC\}/\dg\{\psiTPC\}.
    \label{eqE}
\end{equation}
\end{linenomath}
Note the only two free parameters $a$ and A can be measured experimentally.

\begin{figure}[htbp!]
    \centering
    \includegraphics[width=6.0cm]{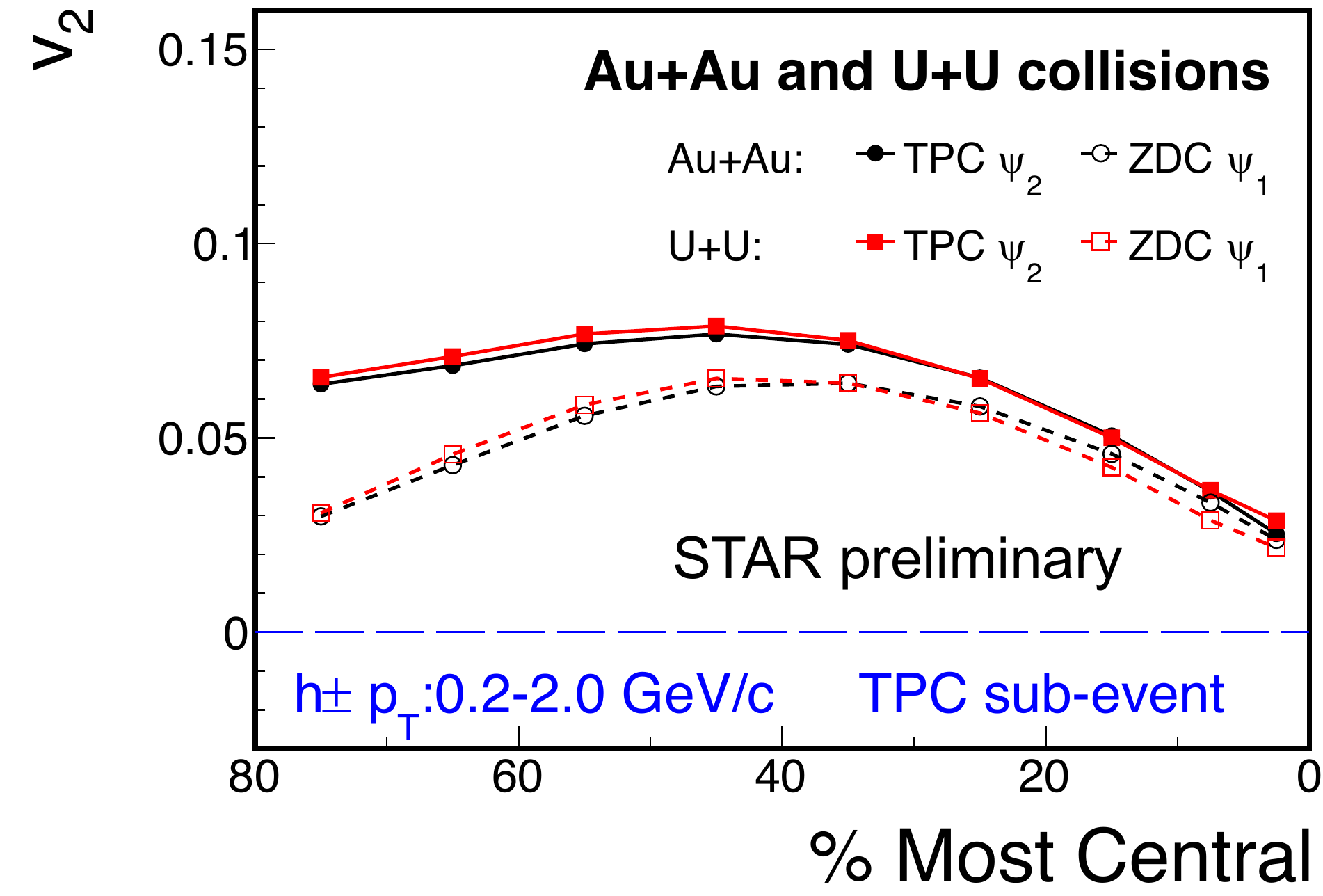}
    \includegraphics[width=6.0cm]{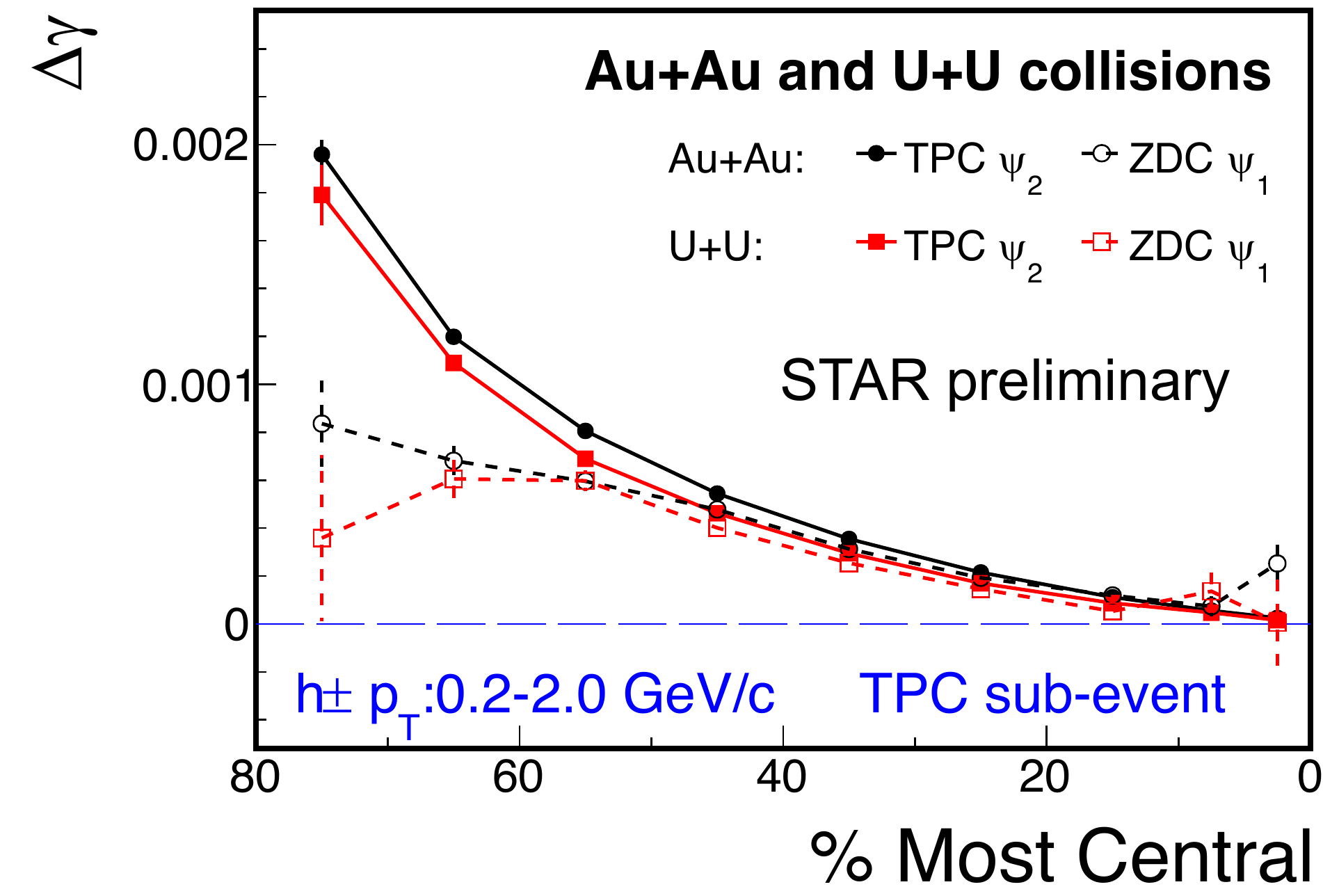}
    \includegraphics[width=6.0cm]{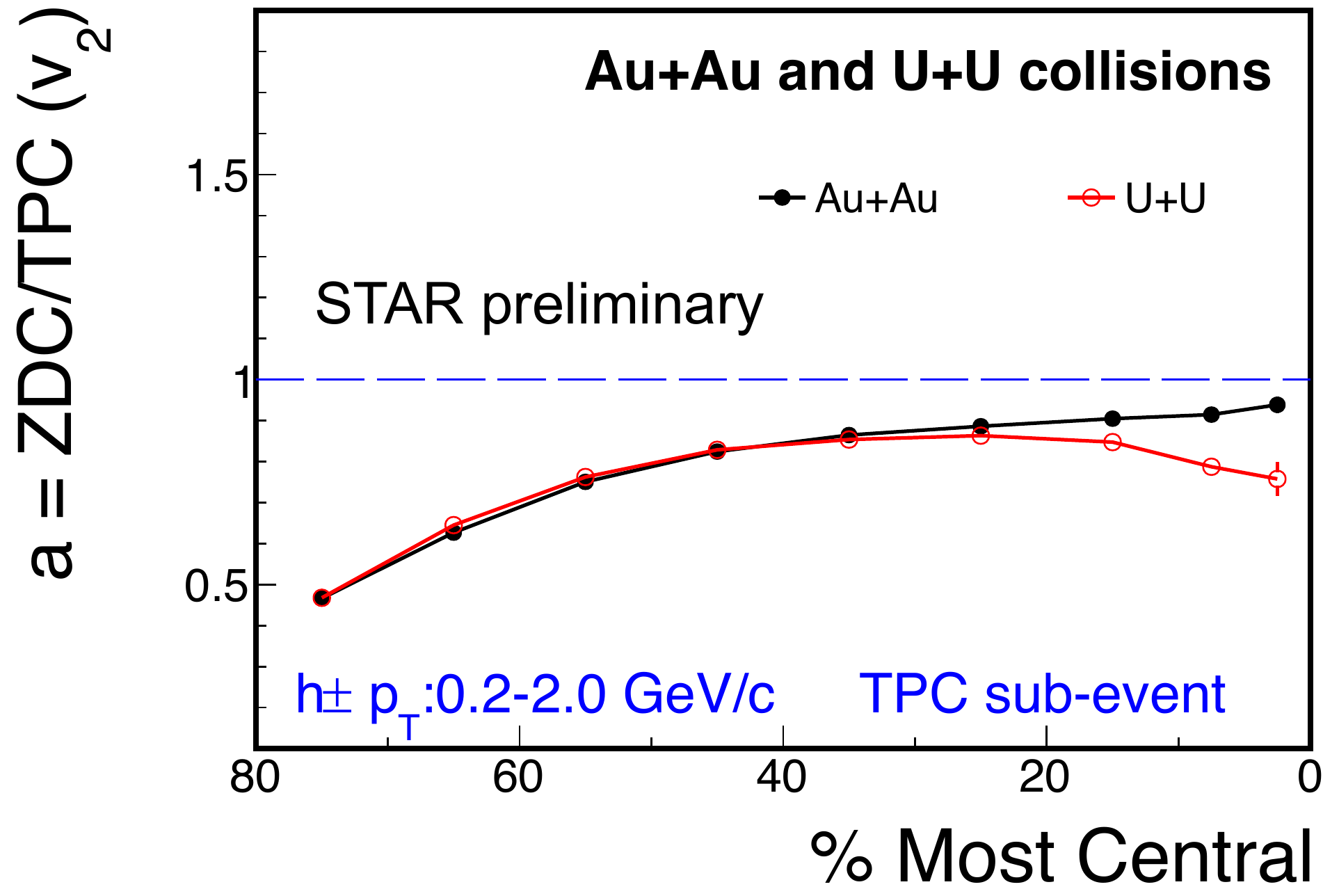}
    \includegraphics[width=6.0cm]{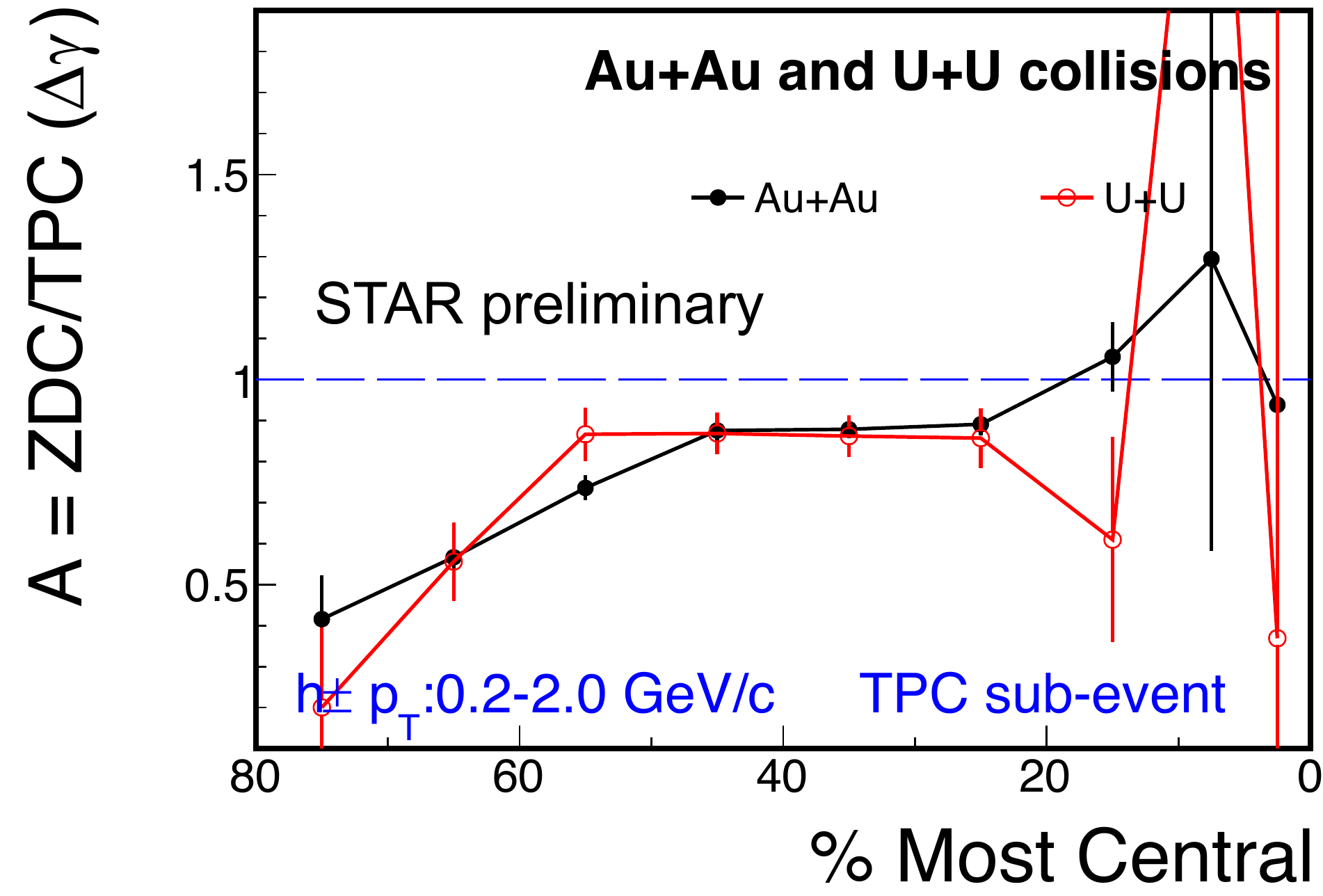}
    \caption{
		The centrality dependence of the $v_2$ (top left) and $\dg$ (top right) measured with respect to the ZDC and TPC event planes.
        The corresponding ratios of the $v_2$ (bottom left) and $\dg$ (bottom right) measured with respect to these two planes. 
    }
    \label{fig2}
\end{figure}

Applying this method, we have previously reported the measurements of possible CME signal fraction in 200 GeV Au+Au collisions, revealing dominant background contribution~\cite{Zhao:2018blc}. 
Here, we report our findings in U+U collisions where the spectator-participant plane correlations are expected to differ from those in Au+Au collisions.
Top panels in Fig.~\ref{fig2} show the measured $v_{2}$ (left) and $\dg$ (right)
with respect to the $\psiZDC$ and $\psiTPC$,
as a functions of collision centrality.
Bottom panels in Fig.~\ref{fig2} show the ratio of $v_{2}$ (left)
measured with respect to the $\psiZDC$ and that with respect to the $\psiTPC$, the $a$ in Eq.~\ref{eqC},
and the ratio of $\dg$ (right), the $A$ in Eq.~\ref{eqE},
as functions of the collision centrality in Au+Au 200 GeV and U+U 193 GeV.
Data indicate difference in $v_2$ between central U+U and Au+Au. 
And the ``a" and ``A" are similar both in trend and magnitude, which indicate background contribution dominates in the $\Delta\gamma_{112}$ measurements.

\begin{SCfigure}
    \centering
    \includegraphics[width=6.2cm]{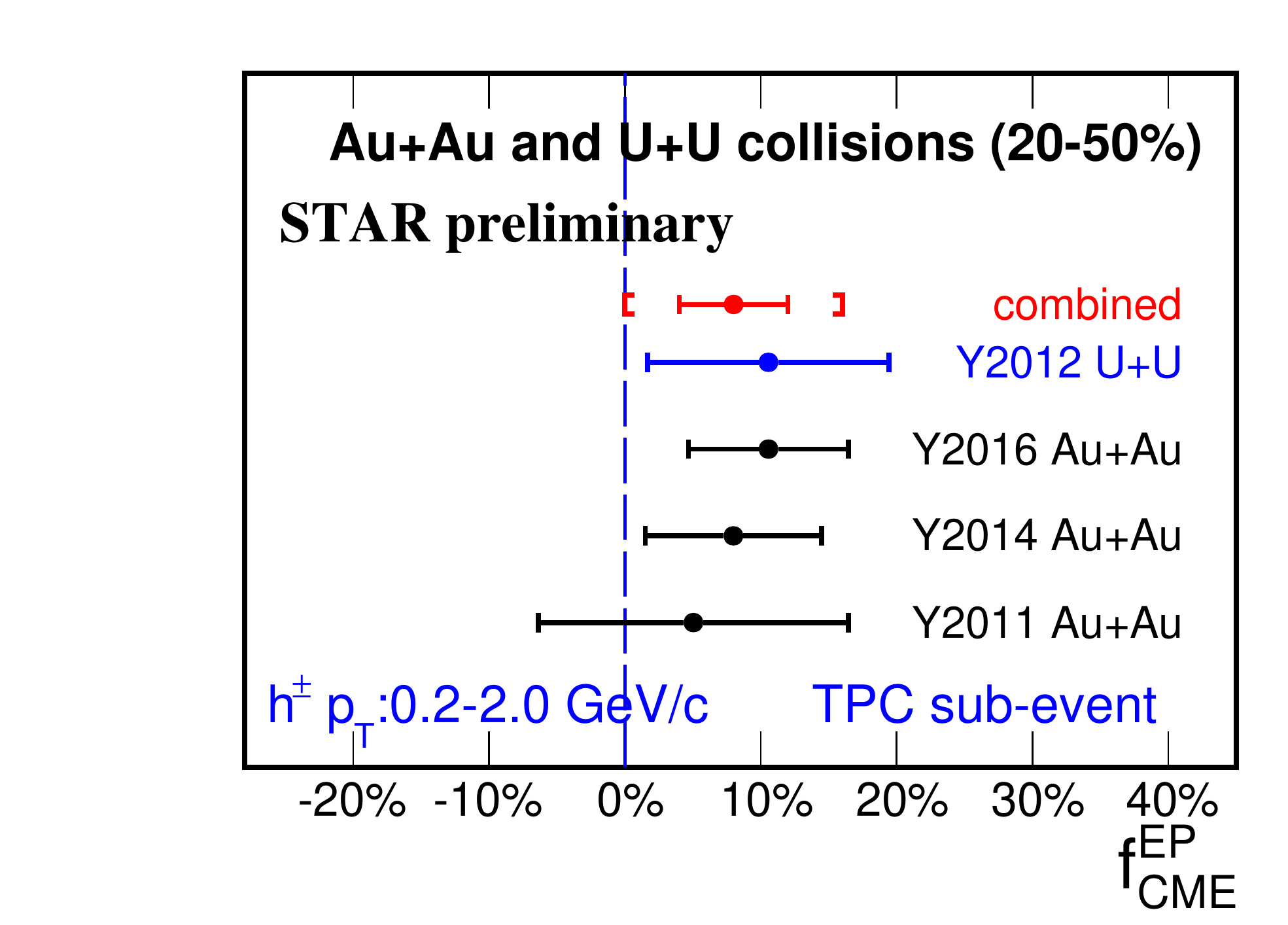}
    \caption{
		The extracted CME fractions ($f_{\rm CME}$) from Au+Au 200 GeV and U+U 193 GeV.
    }
    \label{fig3}
\end{SCfigure}

Figure~\ref{fig3} shows the extracted CME fractions ($f_{\rm CME}$) at Au+Au 200 GeV and U+U 193 GeV. The combined result is $f_{\rm CME}=8\pm4\pm8\%$.

\begin{SCfigure}
    \centering
    \includegraphics[width=5.8cm]{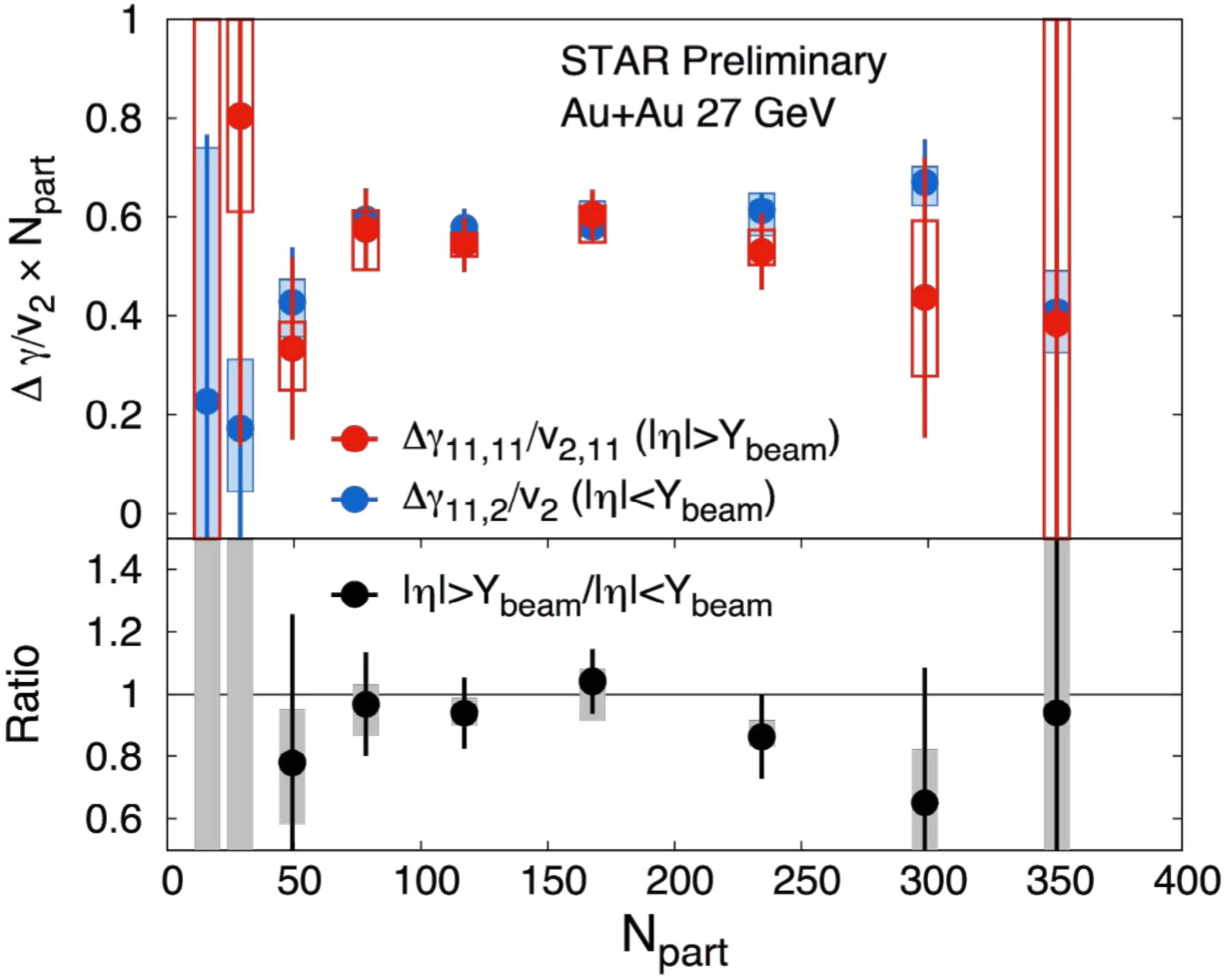}
    \caption{
		(Top) The multiplicity and $v_{2}$ scaled $\dg$ measured with respect to the participant and spectator planes from EPD. 
		(Bottom) The corresponding ratio of the scaled $\dg$ between participant and spectator planes. 
    }
    \label{fig4}
\end{SCfigure}

In Au+Au collisions at $\sqrt{s_{NN}}$=27 GeV, the differential $\dg$ measurements can be achieved by the newly installed Event Plane Detector (EPD) ($2.1<|\eta|<5.1$)~\cite{Adams:2019fpo}.
At this energy, the rapidity of the colliding beam ($y_{beam}$=3.4) falls 
in the middle of EPD acceptance. 
Therefore, the EPD can provide an unique way to search for CME using both $\psiPP$, by outer EPD, and $\psiSP$, by inner EPD. 
Top panel in Fig.~\ref{fig4} shows the multiplicity and $v_{2}$ scaled $\dg$ measurements with respect to the $\psiPP$ and $\psiSP$ from EPD~\cite{Subikash:2019QM}. 
The bottom panel shows that the corresponding ratio 
of $v_2$ or $\dg$ measurements with $\psiSP$ over the one with $\psiPP$ 
is consistent with unity with large uncertainty, 
indicating CME fraction is consistent with zero.
More quantitative studies are in progress.

\section{Summary}
In summary, we report mixed-harmonic three-particle correlation studies in Au+Au and U+U collisions at $\sqrt{s_{NN}}$=200 and 193 GeV, respectively.
The results indicate that background models capture most of the observed trends.
Meanwhile interesting features are observed in ultra-central Au+Au and U+U collisions, which need further investigations.
We also report $v_2$ and $\dg$ measurements with respect to $\psiZDC$ and $\psiTPC$,
and extract the possible CME signal fraction assuming the proportionality of the CME and background to the projection onto the corresponding plane.
The extracted possible CME fraction is $(8\pm4\pm8)\%$ averaged over 20-50$\%$ centrality in Au+Au 200 GeV and U+U 193 GeV collisions.
We further explore the Au+Au 27 GeV data, where the newly installed EPD can be sensitive to both spectator and participant planes.

\vspace{3 mm}
$\bold{Acknowledgments}$ This work was supported by the U.S. Department of Energy (Grant No. de-sc0012910).

\bibliographystyle{elsarticle-num}
\bibliography{ref}








\end{document}